\documentclass[conference]{IEEEtran}
\IEEEoverridecommandlockouts

\usepackage[table]{xcolor}
\usepackage{algorithm}
\usepackage{amsmath,amssymb,amsfonts}
\usepackage{algorithmic}
\usepackage{graphicx}
\usepackage{textcomp}
\usepackage{flushend}
\usepackage{xcolor}
\usepackage{subfig}
\usepackage[numbers,sort&compress]{natbib}

\def\BibTeX{{\rm B\kern-.05em{\sc i\kern-.025em b}\kern-.08em
    T\kern-.1667em\lower.7ex\hbox{E}\kern-.125emX}}

\begin{document}
\captionsetup[figure]{labelsep=period}

\author{
	\IEEEauthorblockN{Di Cui, Siqi Wang, Yong Luo, Xingyu Li, Jie Dai, Lu Wang, Qingshan Li*\thanks{*Corresponding author.}}
	\IEEEauthorblockA{School of Computer Science and Technology, Xidian University, Xi’an 710049, China} 
	\IEEEauthorblockA{cuidi@xidian.edu.cn; qshli@mail.xidian.edu.cn}
}

\title{RMove: Recommending Move Method Refactoring Opportunities using Structural and Semantic Representations of Code}

\maketitle

\begin{abstract}
Incorrect placement of methods within classes is a typical code smell called Feature Envy, which causes additional maintenance and cost during evolution. To remove this design flaw, several Move Method refactoring tools have been proposed. To the best of our knowledge, state-of-the-art related techniques can be broadly divided into two categories: the first line is non-machine-learning-based approaches built on software measurement, while the selection and thresholds of software metrics heavily rely on expert knowledge. The second line is machine learning-based approaches, which suggest Move Method refactoring by learning to extract features from code information. However, most approaches in this line treat different forms of code information identically, disregarding their significant variation on data analysis. In this paper, we propose an approach to recommend Move Method refactoring named RMove by automatically learning structural and semantic representation from code fragment respectively. We concatenate these representations together and further train the machine learning classifiers to guide the movement of method to suitable classes. We evaluate our approach on two publicly available datasets. The results show that our approach outperforms three state-of-the-art refactoring tools including PathMove, JDeodorant, and JMove in effectiveness and usefulness. The results also unveil useful findings and provide new insights that benefit other types of feature envy refactoring techniques.
\end{abstract}


\section{Introduction}


For a long time, the significance of architectural design decisions has been acknowledged in the software research and industry community. The placement of methods within classes in an object-oriented system is a critical criterion for software architecture maintenance. During the software evolution, however, developers may inadvertently and unintentionally implement methods in inappropriate classes, resulting in a typical code smell: Feature Envy \cite{fowler2018refactoring}. Previous studies \cite{d2010impact,sjoberg2012quantifying} also indicated that the Feature Envy is one of the most recurring code smells, negatively and seriously affecting the software system's maintainability. 

To tackle this design flaw, several automatic Move Method refactoring approaches have been proposed, which moves the inappropriate method from its current class to its enviable class. This code transformation eliminates the Feature Envy by improving the internal code structure without changing its external behaviours. Most of these approaches can be more broadly divided into two categories: The first line is metric-based approaches, built on software measurement, such as cohesion and coupling. Although these approaches can intuitively characterize Move Method refactoring from the structural and semantic perspective, the selection and thresholds of software metrics heavily rely on expert knowledge. The second line is machine learning-based approaches, which suggest Move Method refactoring by learning to extract features from the source code. However, in most cases, these approaches treat different forms of code information identically, disregarding their significant variation in data analysis. In reality, various forms of code information, such as structural information and semantic information, require drastically different machine learning algorithms for extracting features. 


In this paper, we combined the comprehensive analysis of metric-based approaches and the automatic feature extraction of machine learning-based approach, and proposed an approach to recommend Move Method refactoring named RMove by learning structural and semantic representation of code fragment separately. 

To capture the structural representation of code, we are motivated by the work of Qu et al. \cite{qu2021evaluating}. Their results demonstrated that the graph embedding technique: node2vec \cite{grover2016node2vec} can effectively characterize the topology of code structure and encode them into low-dimensional vector space as the structural representation of code. Their results presented that these extracted representations are proved to be practical in predicting bugs. Thus, in our method refactoring recommendation task, followed by the work of Qu et al. \cite{qu2021evaluating}, we collect method dependency network as structural information. We further investigate 7 graph embedding techniques to capture structural representations of code based on collected data and make a systematic comparison.


To capture the semantic representation of code, we are motivated by the work of Alon et al. \cite{alon2019code2vec, alon2018codeseq}. Different from the previous techniques capturing program semantics from identifiers and comments using bag-of-words model \cite{bavota2013empirical,cui2019investigating,cui2022towards}, they use code embedding techniques learn continuous distributed vectors from AST paths using graph neural network as semantic representation, mapping semantically similar code snippets to close vectors. Their results also demonstrated that these extracted representations are proved to be useful in predicting method names. Therefore, in our method refactoring recommendation task, followed by the work of Alon et al. \cite{alon2019code2vec, alon2018codeseq}, we collect AST path as semantic information. We employ 2 code embedding techniques to capture semantic representations of code based on collected data and further investigate their impact on the recommendation's performance.


The procedure of our approach: RMove is demonstrated as follows: We first extract method structural and semantic information from the dataset. Next, we create the structural and semantic representation of collected data. {We further normalized and fuse them together as the hybrid representation}. Finally, based on these hybrid representations, we train machine learning classifier to suggest moving a target method to a more structurally and semantically similar class. We evaluate our approach using two publicly available datasets including a synthetic dataset of injected instances \cite{novozhilov2019evaluation} and a real-world dataset of instances annotated by experts \cite{terra2018jmove}. In terms of accuracy and effectiveness, we make a systematic comparison of our approach and other state-of-the-art tools. The results suggest that our approach outperforms state-of-the-art tools such as PathMove \cite{kurbatova2020recommendation}, JDeodorant \cite{tsantalis2009identification}, and JMove \cite{terra2018jmove}. The results also unveil useful findings and provide new insights that benefit other types of feature envy refactoring techniques like move field, move class and move package.

In summary, we make the following contributions:
\begin{itemize}
    \item A new perspective to recommend Move Method refactoring opportunities by exploiting structural and semantic representations of code snippets. 
    \item A systematic exploration of implementations of our approach based on combinations of 2 code embedding techniques, 7 graph embedding techniques, and 9 machine learning classifiers. The results suggest that  Code2Vec+SDNE (CV+SN), Code2Seq+Line (CS+LN), and Code2Seq+SDNE (CS+SN) achieve the best results.
    \item A comprehensive evaluation of our approach on the publicly available dataset. Our approach demonstrates an increase of 14\%-36\% in precision, 19\%-45\% in recall, and 27\%-44\% in f1-measure compared to stat-of-the-art tools: PathMove, JDeodorant, and JMove. 
    \item A benchmark to investigate the effectiveness of structural and semantic representation of code snippets on two widely used datasets. All data are publically available \cite{data}.
\end{itemize}


\section{Preliminary}
\label{sec:preliminary}

In this section, we explain the terminologies used in our paper.

\textbf{Move Method Refactoring Detection.} The Move Method refactoring detection can be regarded as discovering a set of movable methods and target classes from the source code, which is defined as $MoveMethodSet$. Each item in $MoveMethodSet$ can be further modeled as a set of three elements, which is formally defined as follows: 
\begin{equation}
    MoveMethodSet = \{(m_i,sc_i,tc_i) \; | \; i=1,2, \cdots, k\}
\end{equation}
where $m_i$ represents the potentially movable method. $sc_i$ represents the source class which $m_i$ belongs to. $tc_i$ represents the corresponding target class for $m_i$.

\begin{figure}[htbp]
\centerline{\includegraphics[scale=.4]{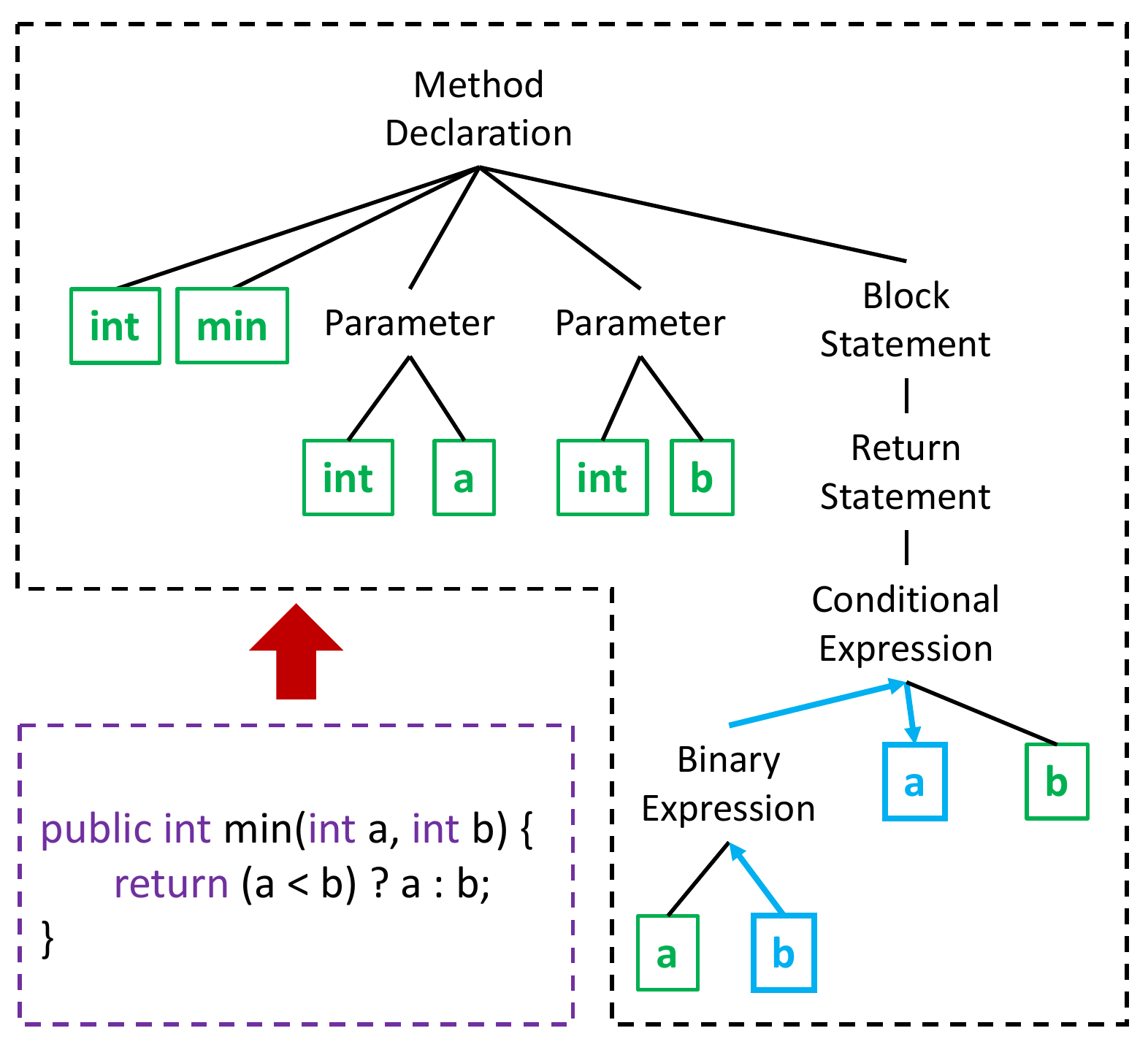}}
\caption{A illustrated example of extracted method semantic information}
\label{fig.semanticinfo}
\end{figure}

To diagnose whether a method is movable, state-of-the-art related techniques analyze its structural and semantic information from the code snippets, which are further illustrated as follows:

\textbf{Method Semantic Information.} For each method, we first parse involved the code snippets into the Abstract Syntax Tree: $AST$, and further iteratively extract the path between leaf nodes from the parsed Abstract Syntax Tree as Method Semantic Information, which is formally defined as $PathSet$: 
\begin{equation}
    PathSet = \{(n_i,n_j, path(n_i,n_j)) \; | \; n_i, n_j \in AST \}
\end{equation}
where $n_i$ and $n_j$ represent a pair of leaf nodes in Abstract Syntax Tree: $AST$. $path(n_i,n_j)$ represents the path between $n_i$ and $n_j$, composed up of a sequence of intermediate AST nodes, which is obtained by traversing through their lowest common ancestor. Fig.~\ref{fig.semanticinfo} illustrates related concepts, including a fragment of code snippet, its corresponding parsed AST, and a path between two leaf nodes highlighted in blue. Arrows in blue in Fig.~\ref{fig.semanticinfo} present the path between two AST leaf nodes: $b$ and $a$, which is represented as $Path(b,a)$: $\{ b \uparrow  BinaryExpression \uparrow ConditionalExpression \downarrow a \}$. Furthermore, all paths: $PathSet$ are gathered to obtain Method Semantic Information.

\begin{figure}[htbp]
\centerline{\includegraphics[scale=.4]{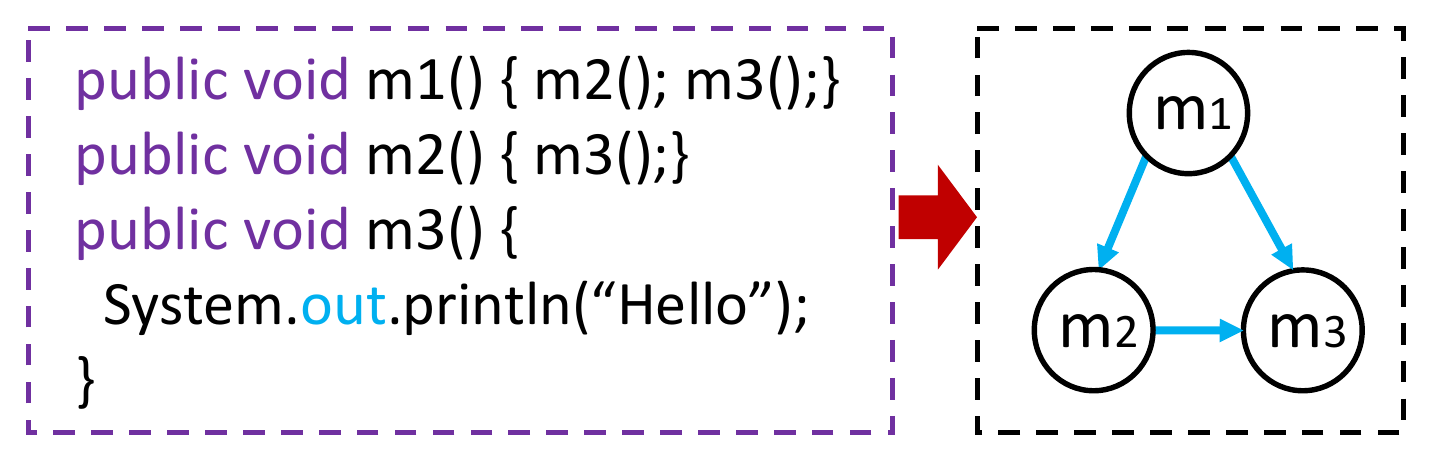}}
\caption{A illustrated example of extracted method structural information}
\label{fig.structureinfo}
\end{figure}

\textbf{Method Structural Information.} Given a method, we first extract Method Dependency Graph (MDG) from the source code and further analyze the topological structure for this method to obtain Method Structural Information. Method Dependency Graph is defined as: 
\begin{equation}
    MDG = \{V, E\}
\end{equation}
where each node $v \in V$ represents a method and the edge $e \in E$ represents the method call dependency relationships. For a pair of methods: $v_i$ and $v_j$, $(v_i,v_j) \in E$ if and only if $v_i$ has at least one method call dependency relationship with $v_j$. Fig.~\ref{fig.structureinfo} illustrates related concepts, including a fragment of code snippet and its corresponding Method Dependency Graph (MDG). These dependency relations: $\{(m_1, m_2), (m_1, m_3), (m_2, m_3)\}$ are presented.

\section{Methodology}
\label{sec:method}

\begin{figure*}[htbp]
\centerline{\includegraphics[scale=.6]{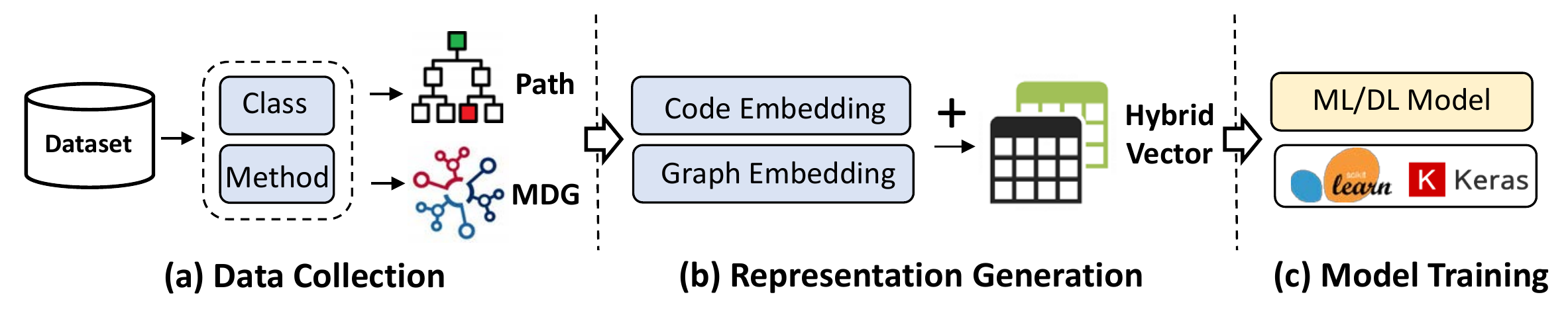}}
\caption{Overview of the proposed approach RMove including data collection, representation generation, and model training.}
\label{fig.overview}
\end{figure*}
\setlength{\belowcaptionskip}{-0.3cm}
Fig.~\ref{fig.overview} presents the overview of our proposed approach: RMove. We implement the automatic Move Method refactoring recommendation system by: 1) mining dataset to gather the Abstract Syntax Tree (AST) and the Method Dependency Graph (MDG), 2) generating hybrid representations with code embedding and graph embedding, and 3) training classification model using machine learning techniques and deep learning techniques. At last, we suggest Move Method refactoring according to the classification result. 

\subsection{Data Collection}

We start with extracting all pairs of movable methods and target classes from the dataset, and then collect method structural and semantic information as follows:

\textbf{Method Semantic Information Collection.} We use \textsc{srcML} \cite{srcml}, a state-of-the-art source code parser, to retrieve the source code of dataset's collected movable methods and target classes. \textsc{srcML} is a lightweight, scalable, multi-language parsing tool for converting source code into the XML format, which supports code exploration, analysis, and manipulation. We then mine AST paths: $PathSet$ with \textsc{astminer} \cite{astminer}, a state-of-the-art static analysis tool, which is applied to the retrieved source code, including movable methods and target classes. \textsc{astminer} is an open-source library for extracting AST paths from the source code. \textsc{astminer} is a efficient, flexible, and extensible tool to support code analysis in various programming languages. We gather all extracted paths:$PathSet$ to obtain Method Semantic Information.

\textbf{Method Structural Information Collection.} we employ \textsc{depends} \cite{depends}, a state-of-the-art static analysis tool, to extract dependencies among methods: $MDG$. \textsc{depends} is a dependency extraction tool for the source code that aims to infer dependency relationships between source code entities, such as files and methods, from various programming languages. \textsc{depends} also provides extensible interfaces to assist with downstream tasks such as architectural analysis and program comprehension. We gather all the Method Dependency Graph: $MDG$ for each involved project in dataset as Method Structural Information.

\subsection{Representation Generation}

For collected method structural and semantic information, we use code embedding techniques and graph embedding techniques to generate its corresponding structural and semantic representations respectively, and further fuse them to generate hybrid representations, which are illustrated as follows: 

\textbf{Code Embedding Generation.} Given each method: $m_i$ and gathered AST path set: $PathSet$, the code embedding: $CE$ is a mapping formally defined as:
\begin{equation}
    CE = \{(m_i, cebd_i) \; | \; path(m_i) \in PathSet \wedge cebd_i \in R^d \} 
\end{equation}
where $path(m_i)$ represents the involved AST paths for $m_i$ and $cebd_i$ represents the code embedding results for $m_i$. In this paper, we explore two state-of-the-art code embedding techniques including Code2Vec and Code2Seq. The basic ideas of these techniques are illustrated as follows:

\textbf{Code2Vec} \cite{alon2019code2vec} is a neural network that automatically generate vectors from code snippets. For AST paths of each method, Code2Vec learns numeric vectors for involved leaf nodes and intermediate nodes respectively. Code2Vec then concatenates these vectors together as a combined context vector. To generate the embedding result for code snippets, based on the attention mechanism, Code2Vec further calculates the weighted average of all combined context vectors by assigning more weights to AST paths with more significant semantics. The obtained fixed-length vectors for code snippets, referred as the code embedding result, can be further used in the downstream tasks. 

\textbf{Code2Seq} \cite{alon2018code2seq} is also a neural network that produces sequences from code snippets. The model tasks the AST path set: $PathSet$ as input and generates distinct embedding results for tokens and paths in the AST path set before combining them into a single vector. To build embedding for tokens, Code2Seq first splits tokens into a sequence of subtokens according to Camel and Snake naming conventions, then transforms subtokens into a sequence of vectors with the embedding matrix: $E^{token}$, and finally combine these vectors to the token embedding. To build embedding for paths, code2seq embeds AST node types into numerical vectors with another embedding matrix $E^{node}$ and thus produce a sequence of vectors. The obtained sequences  then processed through an LSTM with the path's embedding determined by the LSTM's last state. The Code2Seq's final step is to combine the generated vectors into the embedding result. To support this step, Code2Seq concatenates these vectors via a one-layer fully-connected neural network. 

Implementations and hyper-parameter settings of code embedding techniques are introduced in Section~\ref{sec:experiment-setup}, referred to previous papers \cite{alon2019code2vec,alon2018code2seq}. Table~\ref{tab:embedding-example} presents the code embedding results of the method: validateLiteralPresence in the open source project: PMD \cite{pmd}. We observed that these code embedding results are vastly different. 

\begin{table}[]
\small
\caption{The embedding results of the method: validateLiteralPresence in the open source project: PMD}
\begin{center}
\label{tab:embedding-example}
\begin{tabular}{lrrrrr}
\hline
Embedding & \multicolumn{1}{c}{$\textrm{feature}_1$} & \multicolumn{1}{c}{$\textrm{feature}_2$} & \multicolumn{1}{c}{$\textrm{feature}_3$} & \multicolumn{1}{c}{…} & \multicolumn{1}{c}{$\textrm{feature}_\textrm{n}$} \\
\hline
Code2Vec & 0.98 & 0.16 & 0.98 & … & 0.87 \\
Code2Seq & 0.18 & -0.14 & 0.19 & … & -0.06 \\
DeepWalk & -1.25 & 0.62 & 0.19 & … & -0.003 \\
GraRep & 0.18 & -0.14 & 0.19 & … & -0.06 \\
Line & -0.02 & -0.25 & 0.43 & … & -0.06 \\
Node2Vec & -0.07 & 0.01 & -0.31 & … & -0.03 \\
ProNE & -0.004 & -0.31 & 0.38 & … & 0.01 \\
SDNE & 3.94 & 4.57 & 2.76 & … & -0.008 \\
Walklets & -3.16 & -1.37 & -0.44 & … & - 0.04 \\ \hline

\end{tabular}
\end{center}
\end{table}
\setlength{\belowcaptionskip}{-0.1cm}

\textbf{Graph Embedding Generation.} For a Method Dependency Graph $MDG=\{V, E\}$, the graph embedding: $GE$ is a mapping formally defined as: 
\begin{equation}
\setlength{\abovedisplayskip}{5pt}
\setlength{\belowdisplayskip}{5pt}
    GE = \{(m_i, gebd_i) \; | \; m_i \in V \wedge gebd_i \in R^d \wedge d << |V|\} 
\end{equation}
where $m_i$ represents the node in Method Dependency Graph: $MDG$ and $gembedding_i$ represents the graph embedding results for $m_i$. In this paper, we explore seven state-of-the-art graph embedding techniques including DeepWalk, Node2Vec, Walklets,  GraRep, Line, ProNE, and SDNE. The basic ideas of these techniques are illustrated as follows:

\textbf{DeepWalk} \cite{perozzi2014deepwalk} and \textbf{Node2Vec} \cite{grover2016node2vec} employ random walk to construct sample neighborhoods for nodes in graph based on a Skip-gram \cite{mikolov2013efficient} Natural Language Processing (NLP) model. The goal of Skip-gram is to maximize the likelihood of words appearing in a sliding window co-occurring. During random walks, each path sampled from a graph correlates to a sentence from the corpus in NLP, where each node correlates to a word. For these paths, the Skip-gram then is applied to the maximization of the probabilities of having a node's neighborhood based on its embedding results, which implemented with Stochastic Gradient Descent (SGD) and backpropagation on single hidden-layer feedforward neural network \cite{grover2016node2vec}. In comparison to DeepWalk, Node2Vec uses a more flexible notion of the node’s neighborhoods and a more efficient graph searching algorithm achieving the trade-off between Breadth-first Sampling (BFS) and Depth-first Sampling (DFS) \cite{goyal2018graph}.

\textbf{Walklets} \cite{perozzi2017don} is another random walk-based graph embedding technique. In comparison to DeepWalk and Node2Vec, Walklets explicitly encode multi-scale relationships between nodes to produce the multi-scale representations for nodes. Walklets first samples short random walks to extract multi-scale relationships. Furthermore, for each random walk, Walklets skips over steps and constructs latent representations, capturing higher order relationships from the adjacency matrix.

\textbf{GraRep} \cite{cao2015grarep} is a matrix factorization-based graph embedding technique. These techniques construct matrices from connections between nodes and factorize them to produce the embedding result. The following matrices are frequently investigated including node adjacency matrix, node transition probability matrix, Laplacian matrix, etc. GraRep focuses on factorizing the node proximity matrix. The time complexity of GraRep is $\textrm{O}(|\textrm{V}|^3)$, having a scalability issue \cite{goyal2018graph}.

\textbf{Line} \cite{tang2015line} calculates graph embedding results by specifying two functions, one for the first-order node proximity and the other for the second-order node proximity. The line then minimizes the combination of these functions. For each pair of nodes with the first-order proximity, Line defines two joint probability distributions, one for the adjacency matrix and the other for the embedding. The Kullback-Leibler (KL) divergence of these distributions is then minimized. The calculation of the second-order proximity follows a similar pattern \cite{goyal2018graph}. 

\textbf{ProNE} \cite{zhang2019prone} is a fast and scalable graph embedding technique that was recently introduced. ProNE includes two steps: the first is to effectively initialize graph embedding results by phrasing the problem as sparse matrix factorization, motivated by the long-tailed distribution of most graphs and their sparsity. The second stage is to propagate the initial embedding result using the higher-order Cheeger's inequality \cite{lee2014multiway}, aiming at capturing the graph's localized clustering information. The experimental results \cite{zhang2019prone} reveal that ProNE is 10 to 400 times faster than DeepWalk, Node2Vec and Line. ProNE's performance in the multi-label node classification task also outperformed existing graph embedding techniques. 

\textbf{SDNE} \cite{wang2016structural} employs deep auto-encoders to generate embedding results. The objective of an auto-encoder is to reduce the reconstruction error. Multiple extremely non-linear functions are included in both the encoder and the decoder. The encoder converts input data into the representation space, and the decoder converts representation space into the reconstruction space \cite{cai2018comprehensive}. There are two elements to the model: unsupervised and supervised. The first contains an auto-encoder to reconstruct the node's neighborhood and generate its embedding. The second is built on Laplacian Eigenmaps \cite{belkin2001laplacian} and imposes a penalty when similar nodes are mapped with wrong results in the embedding space \cite{goyal2018graph}.   

Implementations and hyper-parameter settings of graph embedding techniques are introduced in Section~\ref{sec:experiment-setup}, referred to two recent survey papers \cite{cai2018comprehensive, goyal2018graph}. Table~\ref{tab:embedding-example} also presents the graph embedding results of the method: validateLiteralPresence in the open source project: PMD \cite{pmd}. We observed that these graph embedding results are drastically different.

\textbf{Representation Fusing.} We first normalize code embedding and graph embedding results respectively. Next, for each method: $m_i$, we concatenate its normalized code embedding and graph embedding as the hybrid embedding: $hebd(m_i)$ with a tuning parameter: $\alpha$, which is formally defined as:
\begin{equation}
    \begin{split}
    hebd(m_i)=[\alpha \times cebd_i,(1-\alpha) \times gebd_i] (m_i, cebd_i) \in NCE \\ \wedge (m_i,gebd_i) \in NGE
    \end{split}
\end{equation}
where $cebd_i$ represents the corresponding code embedding for $m_i$ in the normalized embedding set: $NCE$ and $gebd_i$ represents the corresponding code embedding results for $m_i$ in the normalized embedding set: $NGE$. In our paper, we set the parameter $\alpha$ as 0.5, which means the hybrid embedding is split evenly between the code and graph embedding. 

For each class: $C_j$, its corresponding hybrid embedding is calculated as an element-wise average of hybrid embedding results of contained methods:
\begin{equation}
\setlength{\abovedisplayskip}{4pt}
\setlength{\belowdisplayskip}{4pt}
    hebd(C_j)=\frac{1}{|C_j|}\sum_{m_i \; \in \; C_j}{hebd(m_i)} 
\end{equation}
where $|C_j|$ represents the number of methods in class: $C_j$.

These fused embedding results for methods and classes are further employed in the model training process.
\vspace{1mm}
\subsection{Model Training}
\vspace{1mm}
For fused hybrid representations, we further generate training data and feed them with various classifiers to build the move method refactoring recommendation system, which is illustrated as follows:

\textbf{Training Data Generation.} We generate training data from a small set of detected move method detection results. Algorithm 1 presents the procedure of training data generation. The input of this algorithm is a set of detected move method results: $MoveMethodSet$. This algorithm inspects each item in $MoveMethodSet$ iteratively. For each item, Line 4 retrieves movable method, source class and target class as $m_i$, $sc_i$, $tc_i$ respectively. Line 6 concatenates hybrid embedding results of $m_i$ and $sc_i$, and assigns with the false label as the negative sample. Line 8 concatenates hybrid embedding results of $m_i$ and $tc_i$, and assigns with the true label as the positive sample. For example, for an instance of detected results concluding a movable method: $a$ from the source class: $A$ can be moved to the target class: $B$, we add a positive sample: ``$m$ should be moved to $B$'' and a negative sample: ``$m$ should not be moved to $A$''. At the same time, if detected results also contain an instance that the method: $m$ from the source class: $A$ can be moved to a target class: $C$, we will additionally add a positive sample: ``$m$ should be moved to $C$'' and a duplicate negative sample: ``$m$ should be moved to $A$'' to balance the dataset.   


    \begin{algorithm}[]
        \caption{AutoTrainingDataGeneration$(MoveMethodSet)$}
        \begin{algorithmic}[1]

        \STATE $TrainData \leftarrow \emptyset \;$ \% initialization
 		\FOR{$item$ in $\textit{MoveMethodSet}$}
		\STATE \% getting method, source class and target class respectively
 		\STATE $m_i, sc_i, tc_i \leftarrow item[0], item[1], item[2] \;$ 
 		\STATE \% adding negative samples
 		\STATE $TrainData.add(concat(hebd(m_i), hebd(sc_i)), False)$ 
 		\STATE \% adding positive samples
		\STATE $TrainData.add(concat(hebd(m_i), hebd(tc_i)), True)$ 
  		\ENDFOR  
        \end{algorithmic}
    \end{algorithm}
\vspace{-5pt}

\vspace{1.5mm}
\textbf{Classifier Selection.} Machine learning and deep learning models are frequently investigated for software data classification, which acquire classification knowledge through intensive training. In this paper, we employ 6 machine learning models and 3 deep learning models including Decision Tree (DT), Naive Bayes (NB), Support Vector Machine (SVM), Logistic Regression (LR), Random Forest (RF), Extreme Gradient Boosting (XGB), Convolutional Neural Network (CNN), Long Short Term Memory Recurrent Neural Network (LSTM), and Gated Recurrent Units Recurrent Neural Network (GRU). 

Decision Tree is a flowchart-like structural classifier frequently employed in fast data analysis tasks \cite{safavian1991survey}. Naive Bayes is a classifier according to Bayes’ Theorem with the independent assumption of predictors \cite{rish2001empirical}. Support Vector Machine is a supervised machine learning technique supporting both classification or regression tasks \cite{noble2006support}. Logistic regression is a classifier that estimates probabilities between the categorical dependent variable and independent variables with the logistic or sigmoid function \cite{hosmer2013applied}. Random Forest \cite{pal2005random} and Extreme Gradient Boosting \cite{chen2015xgboost} are ensemble models. Deep learning models, inspired by the human brain structure, have lately acquired enough attention in data classification. We employ widely used models in this field including Convolutional Neural Network (CNN) \cite{girshick2015fast} and two variants of Recurrent Neural Network (RNN) \cite{cho2014learning}: RNN-LSTM, and RNN-GRU.

\section{Evaluation}
\label{sec:evaluation}

We design the evaluation to answer the following three research questions.

\begin{enumerate}
    \item [\textbf{RQ1:}] \textbf{Which embeddings are the most effective in recommending move method refactoring?} The answer to this question would help us better understand the impact of different embedding combinations on RMove's performance.
    \item [\textbf{RQ2:}] \textbf{How accurate is RMove in recommending move method refactoring?} The answer to this question would demonstrate the RMove's performance compared to state-of-the-art tools.
    \item [\textbf{RQ3:}] \textbf{How useful is RMove in recommending move method refactoring?} The answer to this question would shed light on the RMove's effectiveness in practice.
\end{enumerate}

\begin{table}[htbp]
\centering
\small
\caption{Synthetic Dataset's Information}
\label{tab:synthetic}
\resizebox{0.99\columnwidth}{!}{
\begin{tabular}{lccccc}
\hline
Subjects & \multicolumn{1}{c}{\#Version} & \multicolumn{1}{c}{\#LOC} & \multicolumn{1}{c}{\#Classes} & \multicolumn{1}{c}{\#Methods} & \multicolumn{1}{c}{\#MMethods} \\ 
\hline
PMD & 6.13.0 & 119,430 & 1,147 & 8,637 & 127 \\
Cayenne & 4.2 & 275,450 & 1,499 & 12,164 & 93 \\
Pinpoint & 1.9.0 & 290,974 & 2,551 & 17,024 & 105 \\
Jenkins & 1.51 & 155,667 & 768 & 6,292 & 38 \\
Drools & 7.22.0 & 680,234 & 2.758 & 27,793 & 256 \\ 
\hline
\end{tabular}}
\end{table}
\setlength{\belowcaptionskip}{-0.1cm}

\begin{table}[]
\centering
\small
\caption{Real-world Dataset's Information}
\label{tab:realworld}
\resizebox{0.99\columnwidth}{!}{
\begin{tabular}{lccccc}
\hline
Subjects & \#Version & \multicolumn{1}{l}{\#LOC} & \multicolumn{1}{c}{\#Classes} & \multicolumn{1}{c}{\#Methods} & \#MMethods \\ 
\hline
Weka & 3.6.9 & 257,897 & 908 & 16,034 & 31 \\
Ant & 1.8.2 & 103,402 & 760 & 8,586 & 25 \\
FreeCol & 0.10.3 & 93,605 & 535 & 6,616 & 17 \\
JMeter & 2.5.1 & 81,222 & 682 & 7,392 & 25 \\
FreeMind & 0.9.0 & 53,782 & 368 & 4,074 & 12 \\
JTOpen & 7.8 & 340,752 & 1,450 & 22,143 & 39 \\
DrJava & r5387 & 88,631 & 361 & 4,675 & 18 \\
Maven & 3.0.5 & 71,065 & 154 & 1,568 & 24 \\
\hline
\end{tabular}}
\end{table}

\subsection{Experiment Setup}
\label{sec:experiment-setup}

We illustrate the used dataset, evaluation, and experiment settings as follows:

\textbf{Datasets.} We evaluate RMove using two publicly available datasets: a synthetic dataset \cite{novozhilov2019evaluation} and a real-world dataset \cite{terra2018jmove}. 

\textbf{Synthetic dataset.} We use the synthetic dataset to evaluate the impact of various embedding techniques on RMove's performance. The synthetic dataset contains five open-source projects with high quality including PMD \cite{pmd}, Cayenne \cite{cayenne}, Pinpoint \cite{pinpoint}, Jenkins \cite{jenkins}, and Drools \cite{drools}. The basic information of these subjects is presented in Table~\ref{tab:synthetic}, including the number of lines of source code (\#LOC), number of classes (\#Classes), number of methods (\#Methods), and number of movable methods (\#MMethods).

\textbf{Real-world dataset.} We use the real-world dataset to conduct a comparison of RMove and other state-of-the-art refactoring tools. Each instance in the real-world dataset is manually entered by experts and these data are also frequently investigated in previous work \cite{terra2018jmove, kurbatova2020recommendation}. The real-world dataset contains 8 open-source projects with high quality including Weak \cite{weka}, Ant \cite{ant}, FreeCol \cite{freecol}, JMeter \cite{jmeter}, FreeMind \cite{freemind}, JTOpen \cite{jtopen}, DrJava \cite{drjava}, and Maven \cite{maven}. The basic information of these subjects are also presented in Table~\ref{tab:realworld}, including the number of lines of source code (\#LOC), number of classes (\#Classes), number of methods (\#Methods), and number of movable methods (\#MMethods)

\textbf{Evaluation Metrics.} Following the previous work \cite{kurbatova2020recommendation}, we use three widely-used evaluation metrics including precision, recall, and F1 score, defined as follows:
\begin{equation}
    \textrm{Precision} = \frac{\# \; \textrm{of} \; \textrm{correct} \; \textrm{refactorings}}{\# \; \textrm{of} \; \textrm{recommended} \; \textrm{refactorings}}
\end{equation}
\begin{equation}
    \textrm{Recall} = \frac{\# \; of \; \textrm{correct} \; \textrm{refactorings}}{\# \; \textrm{of} \; \textrm{moved} \; \textrm{methods}}
\end{equation}
\begin{equation}
      \textrm{F1-Measure} = 2 \times \frac{\textrm{Precision} \times \textrm{Recall}}{\textrm{Precision}+\textrm{Recall}}
\end{equation}
We calculate precision as the ratio between the number of correct refactorings and the number of all recommended refactorings. We calculate recall as the ratio between the number of correct refactorings and the number of moved methods. We calculate F1-Measure as the harmonic mean of the precision and recall results.


\begin{table*}[]
\caption{Hyper-parameter settings of code embedding and graph embedding techniques}
\label{tab:setting}
\centering
\small
\resizebox{\textwidth}{!}{
\begin{tabular}{ll}
\hline
\multicolumn{1}{c}{Techniques} & \multicolumn{1}{c}{Hyper-parameter settings} \\ 
\hline
Code2Vec & \begin{tabular}[c]{@{}l@{}}num\_epochs=20, train\_batch\_size=1024, test\_batch\_size=1024,  code\_vector\_size=128, path\_embeddings\_size =128\\ 
token\_embeddings\_size = 128, csv\_buffer\_size=100*1024*1024, default\_embeddings\_size = 128 \end{tabular} \\
Code2Seq & \begin{tabular}[c]{@{}l@{}}num\_epochs=3000, test\_batch\_size=256, batch\_size=256, shuffle\_buffer\_size=10000, max\_path\_length = 9 \\ csv\_buffer\_size=100*1024*1024, max\_contexts = 200, embeddings\_size = 128, decoder\_size = 128\end{tabular} \\
DeepWalk & representation\_size=128, clf\_ratio=0.5, number\_walks=10, walk\_length=80, workers=8, window\_size=10 \\
GraRep & kstep=4 \\
Line & representation\_size=128, order=3, negative\_ratio=5, clf\_ratio=0.5 \\
Node2Vec & p=0.25, q=0.25 \\
ProNE & dimension=128, step=10, theta=0.5, mu=0.2 \\
SDNE & alpha=1e-6, beta=5, nu1=1e-5, nu2=1e-4, batch\_size=200, epoch=100 \\
Walklets & dimensions=128, walk-number=5, walk\_length=80, window\_size=5, workers=4, min\_count=1, p=1.0, q=1.0 \\
\hline
\end{tabular}}
\end{table*}
\setlength{\belowcaptionskip}{-0.2cm}

\textbf{Experiment Settings.} We run the  experiments on a 2.4GHz Intel Xeon-4210R server with 10 logical cores and 128GB of memory. We also follow most of default hyper-parameters of code embedding and graph embedding techniques in previous work \cite{alon2019code2vec,alon2018code2seq,qu2021evaluating}, which is presented in Table~\ref{tab:setting}.

In answering RQ1 (embedding evaluation), we train classifiers with various embedding techniques on the whole synthetic data and evaluate its effectiveness on RMove's performance. We implement machine learning classifiers based on the python library: \textsc{scikit-learn} \cite{scikitlearn} and deep learning classifiers based on the python library: \textsc{Keras} \cite{keras}. We use the grid search strategies to automatically tune the hyper-parameters of classifiers \cite{tantithamthavorn2018impact}. We also repeat the 10-fold cross-validation 10 times (10×10) to reduce the bias caused by the randomness in experiments. The evaluation metrics are also calculated as the average value during these times.

In answering RQ2 (accuracy evaluation), we select the top 3 embedding combinations and related trained models according to the answer to RQ1. Furthermore, we evaluate the accuracy of the most effective models of these embedding combinations on the real-world dataset and compare these results to other state-of-the-art refactoring tools.

In answering RQ3 (usefulness evaluation), we conduct a human study by hiring 15 experienced software engineers to analyze recommended instances of move method refactoring for each refactoring tools in RQ2. All the participants are not the authors of this paper. Furthermore, we randomly select 5 instances from the detection result from each refactoring tool to reduce the complexity of analyzing the move method refactoring and thus maintain the concentration of participants.

\subsection{Experiment Result}

\begin{table*}[]
\small
\caption{The performance of combinations of various embedding techniques on the synthetic data. CV: Code2Vec, CS: Code2Seq, DW: DeepWalk, GR: GraRep, LN: Line, PN: ProNE, SN: SDNE, P: Precision, R: Recall, F1: F1-Measure.}
\label{tab:RQ1-result}
\resizebox{\textwidth}{!}{
\begin{tabular}{llllllllllllllllllllll}
\hline
 & \multicolumn{3}{c}{CV+DW} & \multicolumn{3}{c}{CV+GR} & \multicolumn{3}{c}{CV+LN} & \multicolumn{3}{c}{CV+NV} & \multicolumn{3}{c}{CV+PN} & \multicolumn{3}{c}{\cellcolor{gray!50}CV+SN} & \multicolumn{3}{c}{CV+WL} \\
{Model} & \multicolumn{1}{c}{P} & \multicolumn{1}{c}{R} & \multicolumn{1}{c}{F1} & \multicolumn{1}{c}{P} & \multicolumn{1}{c}{R} & \multicolumn{1}{c}{F1} & \multicolumn{1}{c}{P} & \multicolumn{1}{c}{R} & \multicolumn{1}{c}{F1} & \multicolumn{1}{c}{P} & \multicolumn{1}{c}{R} & \multicolumn{1}{c}{F1} & \multicolumn{1}{c}{P} & \multicolumn{1}{c}{R} & \multicolumn{1}{c}{F1} & \multicolumn{1}{c}{P} & \multicolumn{1}{c}{R} & \multicolumn{1}{c}{F1} & \multicolumn{1}{c}{P} & \multicolumn{1}{c}{R} & \multicolumn{1}{c}{F1} \\  \hline
{DT} & .78 & \cellcolor{gray!50}{$.81^+$} & .79 & .82 & .79 & .80 & .75 & \cellcolor{gray!50}{$.81^+$} & .78 & .84 & .80 & \cellcolor{gray!50}{$.82^+$} & .78 & \cellcolor{gray!50}{$.81^+$} & .80 & .83 & .85 & \cellcolor{gray!50}{$.84^+$} & \cellcolor{gray!50}$.81^+$ & \cellcolor{gray!50}{$.83^+$} & \cellcolor{gray!50}{$.82^+$} \\
{NB} & .55 & .35 & .43 & .68 & .48 & .56 & .71 & .51 & .60 & .67 & .46 & .55 & .66 & .52 & .58 & .63 & .86 & .72 & .58 & .52 & .55 \\
{SVM} & .76 & .74 & .75 & .73 & .70 & .71 & .79 & .78 & .78 & .80 & .75 & .78 & .82 & .76 & .79 & .70 & .87 & .78 & .77 & .79 & .78 \\
{LR} & .74 & .74 & .74 & .71 & .78 & .74 & .77 & .77 & .77 & .77 & .75 & .76 & .79 & .80 & .79 & .75 & .81 & .77 & .77 & .75 & .76 \\
{RF} & .83 & .80 & .81 & .81 & .78 & .79 & \cellcolor{gray!50}{$.84^+$} & \cellcolor{gray!50}{$.81^+$} & \cellcolor{gray!50}{$.82^+$} & \cellcolor{gray!50}{$.85^+$} & .74 & .79 & .83 & .76 & .79 & \cellcolor{gray!50}{$.85^+$} & .80 & .82 & .80 & .67 & .73 \\
{XGB} & \cellcolor{gray!50}{$.85^+$} & \cellcolor{gray!50}{$.81^+$} & \cellcolor{gray!50}{$.83^+$} & \cellcolor{gray!50}{$.87^+$} & \cellcolor{gray!50}{$.81^+$} & \cellcolor{gray!50}{$.84^+$} & .82 & .74 & .78 & .80 & .75 & .78 & \cellcolor{gray!50}{$.84^+$} & .78 & \cellcolor{gray!50}{$.81^+$} & \cellcolor{gray!50}{$.85^+$} & .81 & .83 & .79 & .78 & .78 \\
{CNN} & .62 & .74 & .68 & .61 & .79 & .69 & .66 & .68 & .67 & .54 & \cellcolor{gray!50}{$.87^+$} & .67 & .66 & .80 & .72 & .54 & \cellcolor{gray!50}{$.92^+$} & .68 & .72 & .72 & .72 \\
{LSTM} & .74 & .73 & .73 & .65 & .70 & .67 & .77 & .78 & .77 & .76 & .78 & .77 & .78 & .79 & .78 & .76 & .78 & .77 & .72 & .76 & .74 \\
{GRU} & .73 & .73 & .74 & .69 & .47 & .56 & .77 & .69 & .73 & .79 & .67 & .73 & .80 & .69 & .74 & .80 & .73 & .76 & .78 & .65 & .71 \\
{Avg} & .73 & .72 & .72 & .73 & .70 & .71 & .76 & .73 & .74 & .76 & .73 & .74 & .77 & .74 & .76 & .74 & \cellcolor{gray!50}{$.82^*$} & \cellcolor{gray!50}{$.78^*$} & .75 & .72 & .73 \\ \hline
 & \multicolumn{3}{c}{CS+DW} & \multicolumn{3}{c}{CS+GR} & \multicolumn{3}{c}{\cellcolor{gray!50}{CS+LN}} & \multicolumn{3}{c}{CS+NV} & \multicolumn{3}{c}{CS+PN} & \multicolumn{3}{c}{\cellcolor{gray!50}{CS+SN}} & \multicolumn{3}{c}{CS+WL} \\
{Model} & \multicolumn{1}{c}{P} & \multicolumn{1}{c}{R} & \multicolumn{1}{c}{F1} & \multicolumn{1}{c}{P} & \multicolumn{1}{c}{R} & \multicolumn{1}{c}{F1} & \multicolumn{1}{c}{P} & \multicolumn{1}{c}{R} & \multicolumn{1}{c}{F1} & \multicolumn{1}{c}{P} & \multicolumn{1}{c}{R} & \multicolumn{1}{c}{F1} & \multicolumn{1}{c}{P} & \multicolumn{1}{c}{R} & \multicolumn{1}{c}{F1} & \multicolumn{1}{c}{P} & \multicolumn{1}{c}{R} & \multicolumn{1}{c}{F1} & \multicolumn{1}{c}{P} & \multicolumn{1}{c}{R} & \multicolumn{1}{c}{F1} \\ \hline
{DT} & .76 & \cellcolor{gray!50}{$.75^{+}$} & .75 & .82 & .71 & .76 & .84 & .83 & .83 & .74 & .82 & .78 & .81 & .78 & .80 & .84 & .89 & .87 & .78 & \cellcolor{gray!50}{$.71^{+}$} & \cellcolor{gray!50}{$.74^+$} \\
{NB} & .58 & .40 & .47 & .55 & .33 & .42 & .66 & .77 & .71 & .66 & .41 & .51 & .69 & .73 & .71 & .68 & \cellcolor{gray!50}{$.94^+$} & .79 & .55 & .59 & .57 \\
{SVM} & .73 & .67 & .70 & .82 & .77 & .79 & .82 & .75 & .78 & .79 & .73 & .76 & .80 & .78 & .79 & .74 & .83 & .78 & .75 & .66 & .70 \\
{LR} & .68 & .65 & .66 & .83 & .77 & .80 & .84 & .76 & .80 & .81 & .75 & .78 & .83 & .78 & .80 & .75 & .75 & .75 & .68 & .65 & .66 \\
{RF} & \cellcolor{gray!50}{$.85^+$} & .69 & \cellcolor{gray!50}{$.76^+$} & \cellcolor{gray!50}{$.86^+$} & .75 & .78 & .88 & .81 & .84 & \cellcolor{gray!50}{$.87^+$} & .81 & \cellcolor{gray!50}{$.84^+$} & \cellcolor{gray!50}{$.87^+$} & .78 & .82 & \cellcolor{gray!50}{$.92^+$} & .84 & \cellcolor{gray!50}{$.88^+$} & \cellcolor{gray!50}{$.84^+$} & .66 & .73 \\
{XGB} & .81 & .71 & \cellcolor{gray!50}{$.76^+$} & .83 & \cellcolor{gray!50}{$.83^+$} & \cellcolor{gray!50}{$.83^+$} & \cellcolor{gray!50}{$.89^+$} & .87 & \cellcolor{gray!50}{$.88^+$} & .86 & \cellcolor{gray!50}{$.83^+$} & \cellcolor{gray!50}{$.84^+$} & \cellcolor{gray!50}{$.87^+$} & \cellcolor{gray!50}{$.86^+$} & \cellcolor{gray!50}{$.87^+$} & \cellcolor{gray!50}{$.92^+$} & .85 & \cellcolor{gray!50}{$.88^+$} & .78 & .67 & .72 \\
{CNN} & .73 & .67 & .63 & .65 & .63 & .62 & .54 & \cellcolor{gray!50}{$.96^+$} & .70 & .72 & .67 & .63 & .73 & .69 & .65 & .76 & .68 & .64 & .77 & .68 & .63 \\
{LSTM} & .68 & .61 & .64 & .74 & .63 & .68 & .83 & .76 & .79 & .79 & .72 & .76 & .79 & .78 & .78 & .77 & .81 & .79 & .63 & .57 & .60 \\
{GRU} & .71 & .54 & .61 & .71 & .52 & .60 & .75 & .63 & .68 & .78 & .61 & .68 & .77 & .68 & .72 & .78 & .78 & .78 & .65 & .57 & .61 \\
{Avg} & .72 & .63 & .66 & .76 & .66 & .70 & \cellcolor{gray!50}{$.78^*$} & \cellcolor{gray!50}{$.79^*$} & \cellcolor{gray!50}{$.78^*$} & \cellcolor{gray!50}{$.78^*$} & .71 & .73 & \cellcolor{gray!50}{$.79^*$} & .76 & .77 & \cellcolor{gray!50}{$.80^*$} & \cellcolor{gray!50}{$.82^*$} & \cellcolor{gray!50}{$.80^*$} & .71 & .64 & .66 \\ \hline
\end{tabular}}
\end{table*}
\setlength{\belowcaptionskip}{-0.3cm}

\begin{table*}[]
\centering
\caption{The accuracy of various refactoring tools on the real-world dataset. P: Precision, R: Recall, F1: F1-Measure.}
\label{tab:RQ2-result}
\small
\resizebox{\textwidth}{!}{
\begin{tabular}{lllllllllllllllllll}
\hline
 & \multicolumn{3}{c}{PathMove} & \multicolumn{3}{c}{JDeodorant} & \multicolumn{3}{c}{JMove}  & \multicolumn{3}{c}{\cellcolor{gray!50}{RMove-1}} & \multicolumn{3}{c}{RMove-2} & \multicolumn{3}{c}{RMove-3} \\
 Subjects & \multicolumn{1}{c}{P} & \multicolumn{1}{c}{R} & \multicolumn{1}{c}{F1} & \multicolumn{1}{c}{P} & \multicolumn{1}{c}{R} & \multicolumn{1}{c}{F1} & \multicolumn{1}{c}{P} & \multicolumn{1}{c}{R} & \multicolumn{1}{c}{F1} & \multicolumn{1}{c}{P} & \multicolumn{1}{c}{R} & \multicolumn{1}{c}{F1} & \multicolumn{1}{c}{P} & \multicolumn{1}{c}{R} & \multicolumn{1}{c}{F1} & \multicolumn{1}{c}{P} & \multicolumn{1}{c}{R} & \multicolumn{1}{c}{F1} \\ 
\hline
Weka & .224 & .645 & .332 & .059 & .548 & .107 & .108 & .741 & .189 & .508 & \cellcolor{gray!50}{$.959^+$} & \cellcolor{gray!50}{$.664^+$} & \cellcolor{gray!50}{$.514^+$} & .917 & .659 & .485 & .729 & .583 \\
Ant & .197 & .56 & .292 & .171 & .48 & .252 & .173 & .84 & .287 & .491 & \cellcolor{gray!50}{$.905^+$} & \cellcolor{gray!50}{$.637^+$} & \cellcolor{gray!50}{$.503^+$} & .857 & .634 & .472 & .69 & .561 \\
FreeCol & .048 & .647 & .089 & .03 & .294 & .054 & .074 & .764 & .135 & .483 & .879 & .623 & \cellcolor{gray!50}{$.513^+$} & \cellcolor{gray!50}{$.906^+$} & \cellcolor{gray!50}{$.655^+$} & .506 & .772 & .611 \\
JMeter & .264 & .36 & .305 & .236 & .52 & .325 & .275 & .76 & .404 & .466 & \cellcolor{gray!50}{$.923^+$} & \cellcolor{gray!50}{$.62^+$} & .469 & .847 & .604 & \cellcolor{gray!50}{$.473^+$} & .76 & .583 \\
FreeMind & \cellcolor{gray!50}{$.8^+$} & .333 & .47 & .166 & .583 & .258 & .148 & .666 & .242 & .517 & \cellcolor{gray!50}{$.964^+$} & \cellcolor{gray!50}{$.673^+$} & .524 & .857 & .65 & .484 & .764 & .593 \\
JTOpen & .416 & .512 & .459 & .207 & .447 & .283 & .208 & \cellcolor{gray!50}{$.894^+$} & .337 & .484 & .872 & .623 & .497 & .854 & .628 & \cellcolor{gray!50}{$.521^+$} & .812 & \cellcolor{gray!50}{$.635^+$} \\
DrJava & .428 & .5 & .461 & .128 & .555 & .208 & .128 & .777 & .22 & \cellcolor{gray!50}{$.526^+$} & \cellcolor{gray!50}{$.96^+$} & \cellcolor{gray!50}{$.68^+$} & .5 & .825 & .623 & .518 & .81 & .632 \\
Maven & \cellcolor{gray!50}{$.545^+$} & .541 & .543 & .139 & .25 & .179 & .104 & .375 & .163 & .496 & .853 & .627 & .505 & .771 & .61 & .51 & \cellcolor{gray!50}{$.876^+$} & \cellcolor{gray!50}{$.645^+$} \\
Avg & .365 & .512 & .369 & .142 & .46 & .208 & .152 & .727 & .247 & .496 & \cellcolor{gray!50}{$.914^*$} & \cellcolor{gray!50}{$.643^*$} & \cellcolor{gray!50}{$.503^*$} & .854 & .633 & .496 & .777 & .605 \\ 
\hline
\end{tabular}}
\end{table*}
\setlength{\belowcaptionskip}{-0.1cm}

\begin{figure*} 
	\centering
	\subfloat[Precision of embedding combinations]{
		\includegraphics[scale=0.4]{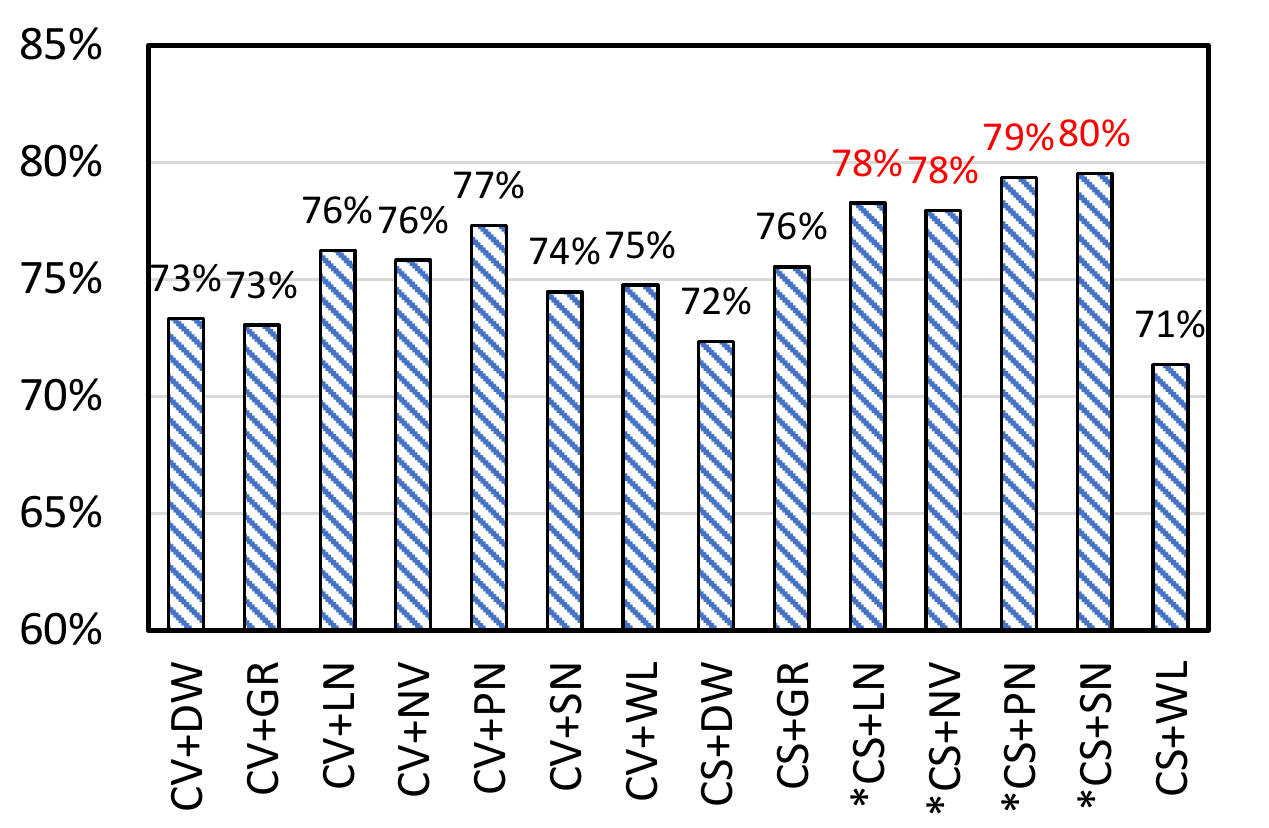}}
	\subfloat[Recall of embedding combinations]{
		\includegraphics[scale=0.4]{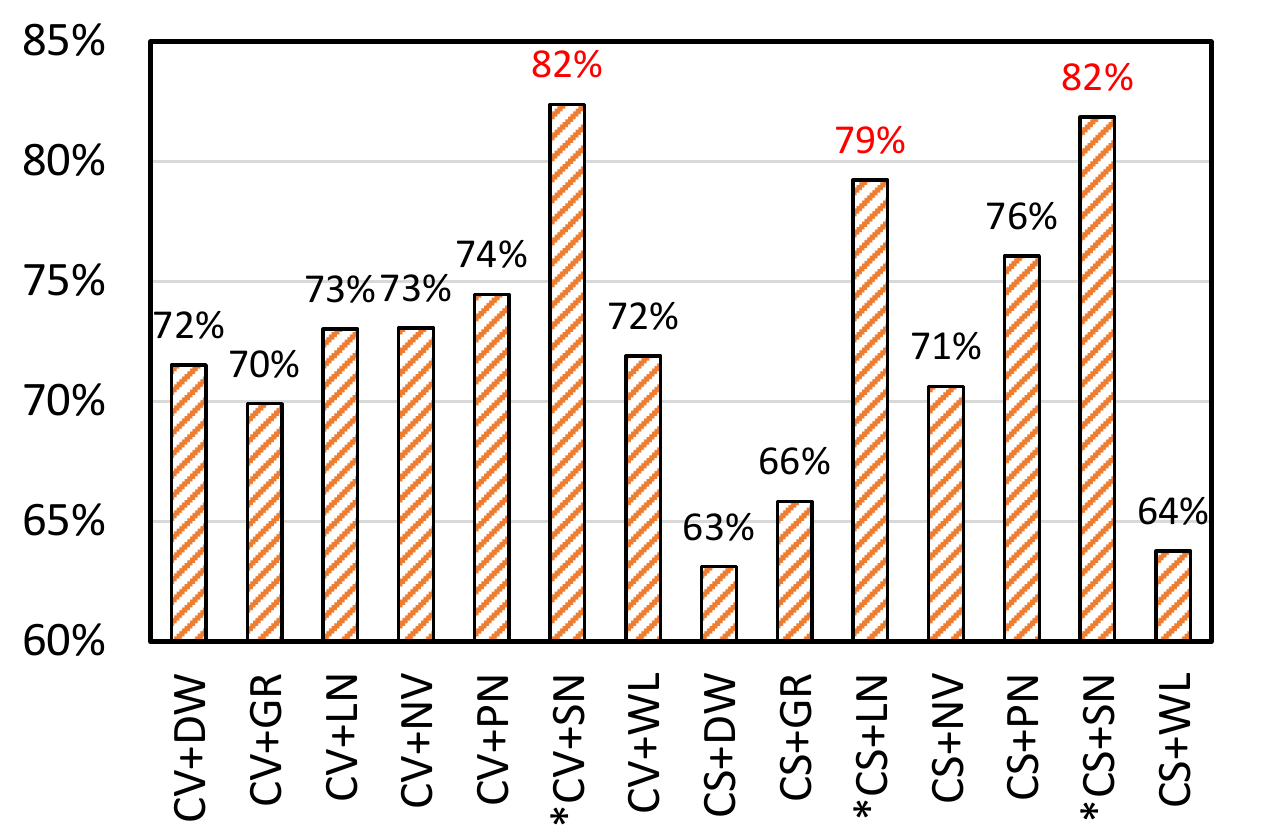}}
	\subfloat[F1-measure of embedding combinations]{
		\includegraphics[scale=0.4]{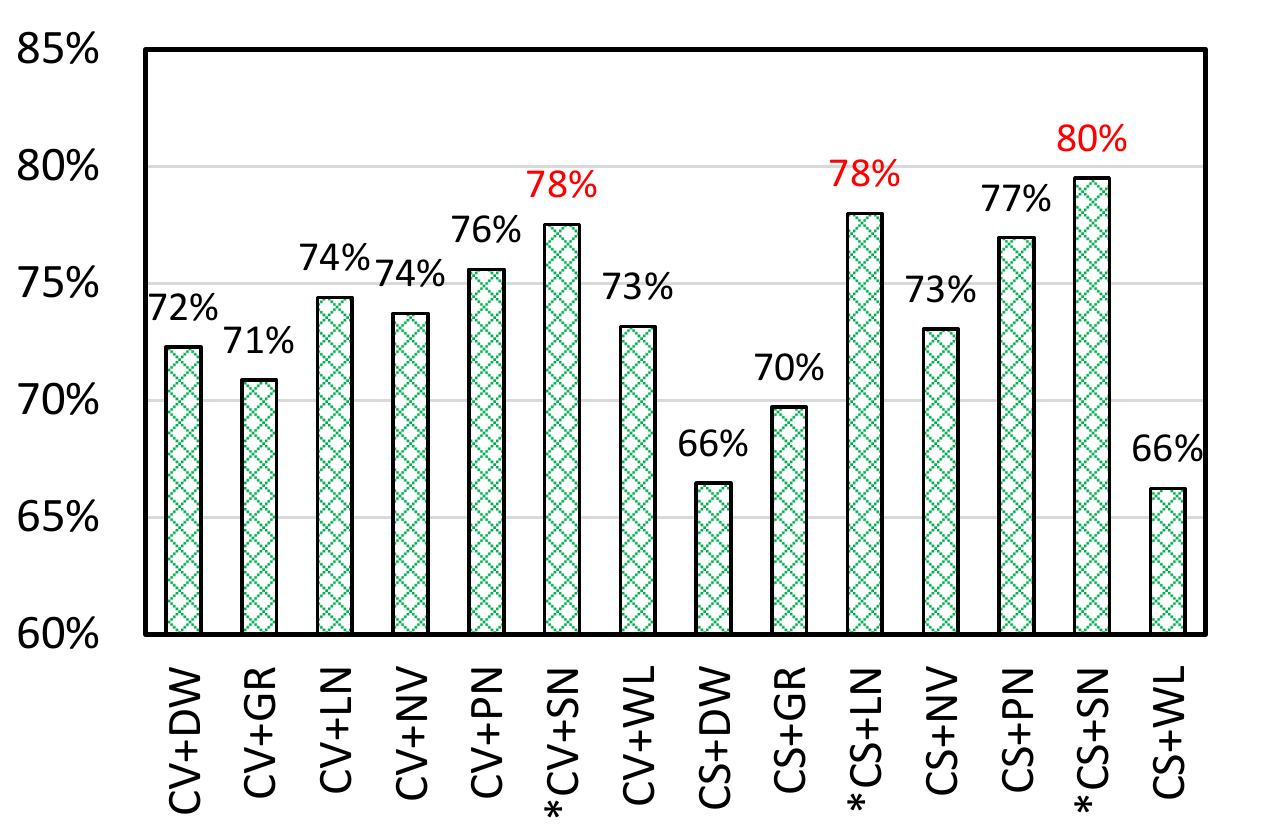}}
	\caption{The average precision, recall, and f1-measure of embedding combinations.}
	\label{fig:RQ1-result}
\end{figure*}

\textbf{Embedding Evaluation (RQ1).} To analyze the effectiveness of various embedding techniques, we conduct a systemic comparison of 14 combinations of 2 code embedding techniques and 7 graph embedding techniques. For each combination, we further investigate its performance with 9 classifiers on the synthetic dataset. Table~\ref{tab:RQ1-result} presents the performance of combinations of various embedding techniques on the synthetic data. For each column, we highlight the greatest precision, recall and f1-measure results with a grey background color and a + mark. For the row: ``avg'', we also highlight the top 3 results of precision, recall, and F1-measure with a grey background color and a * mark. We further perform Kruskal-Wallis test and Dunnett’s test on the experimental results of 14 combinations, which suggests that there is a significant difference among these approaches. Accordingly, we label the top 3 embedding combinations with a grey background color. Fig.~\ref{fig:RQ1-result} further presents the average results of embedding combinations on various classifiers, where the vertical axis shows various embedding combinations, and the horizontal axis shows the precision, recall, and f1-measure respectively. For Fig.~\ref{fig:RQ1-result}.(a)-(c), we label the top 3 results with a red color and corresponding embedding combinations with a * mark.

As presented in Table~\ref{tab:RQ1-result} and Fig.~\ref{fig:RQ1-result}, we observed that 1) Combinations of Code2Vec+SDNE (CV+SN), code2Seq+Line (CS+LN), and Code2Seq+SDNE (CS+SN) outperform other embedding combinations. Code2Seq+SDNE (CS+SN) achieves the greatest results in precision, recall, and f1-measure. Code2Vec+SDNE (CV+SN) presents the greatest recall score and Code2Seq+Line (CS+LN) presents a relatively high precision and f1-measure score. A possible explanation is that SDNE comprehensively captures the method semantic information to improve the recall score whereas Code2Seq efficiently characterizes the method semantic information to increase the precision score. 2) Random Forest (RF) and Extreme Gradient Boosting (XGB) outperform other classifiers in most embedding combinations. Although Decision Tree (DT) performs effectively in specific embedding combinations, Random Forest (RF) and Extreme Gradient Boosting (XGB) present the best precision, recall and f1-measure results in most cases. This result demonstrates that deep learning classifiers do not perform as well as we expected in recommending move method refactorings. The reason might be that software data, unlike image and text, is more suitable for machine learning classifiers.

\noindent\fbox{
 \parbox{0.95\linewidth}{
  \textbf{Answer to RQ1}:
  Code2Vec+SDNE (CV+SN), Code2Seq +Line (CS+LN), and Code2Seq+SDNE (CS+SN) are the most effective embedding combinations.
 }
 }

\begin{figure*} 
	\centering
	\subfloat[Classifier performance on CV+SN]{
		\includegraphics[scale=0.4]{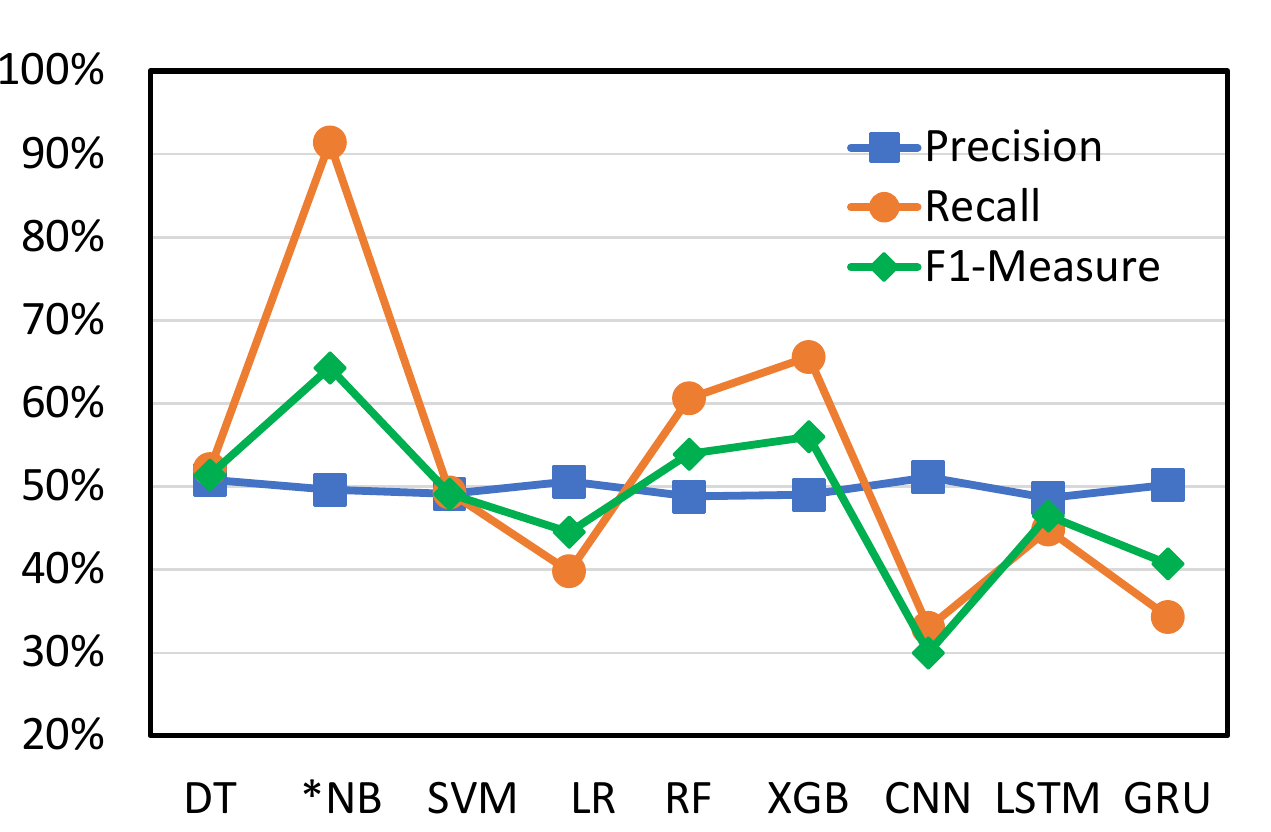}}
	\subfloat[Classifier Performance on CS+LN]{
		\includegraphics[scale=0.4]{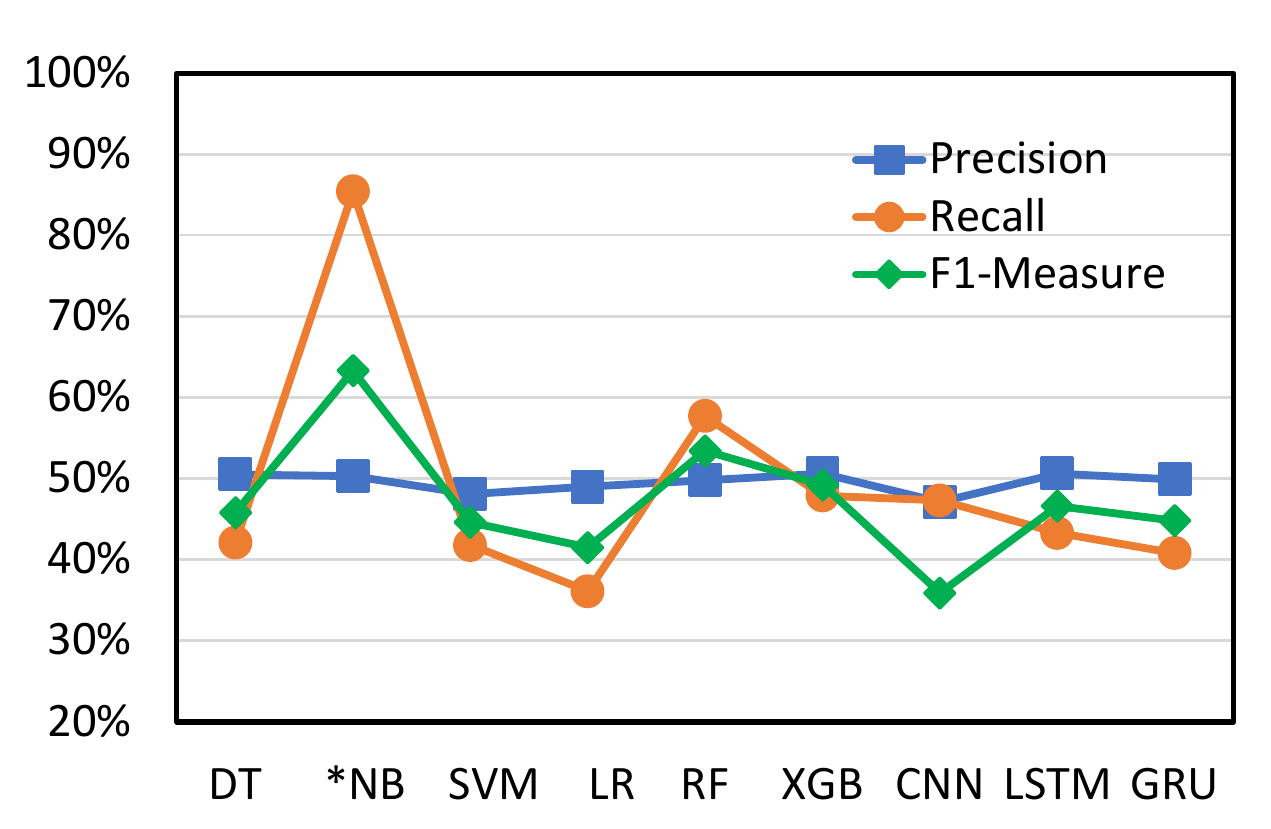}}
	\subfloat[Classifier performance on CS+SN]{
		\includegraphics[scale=0.4]{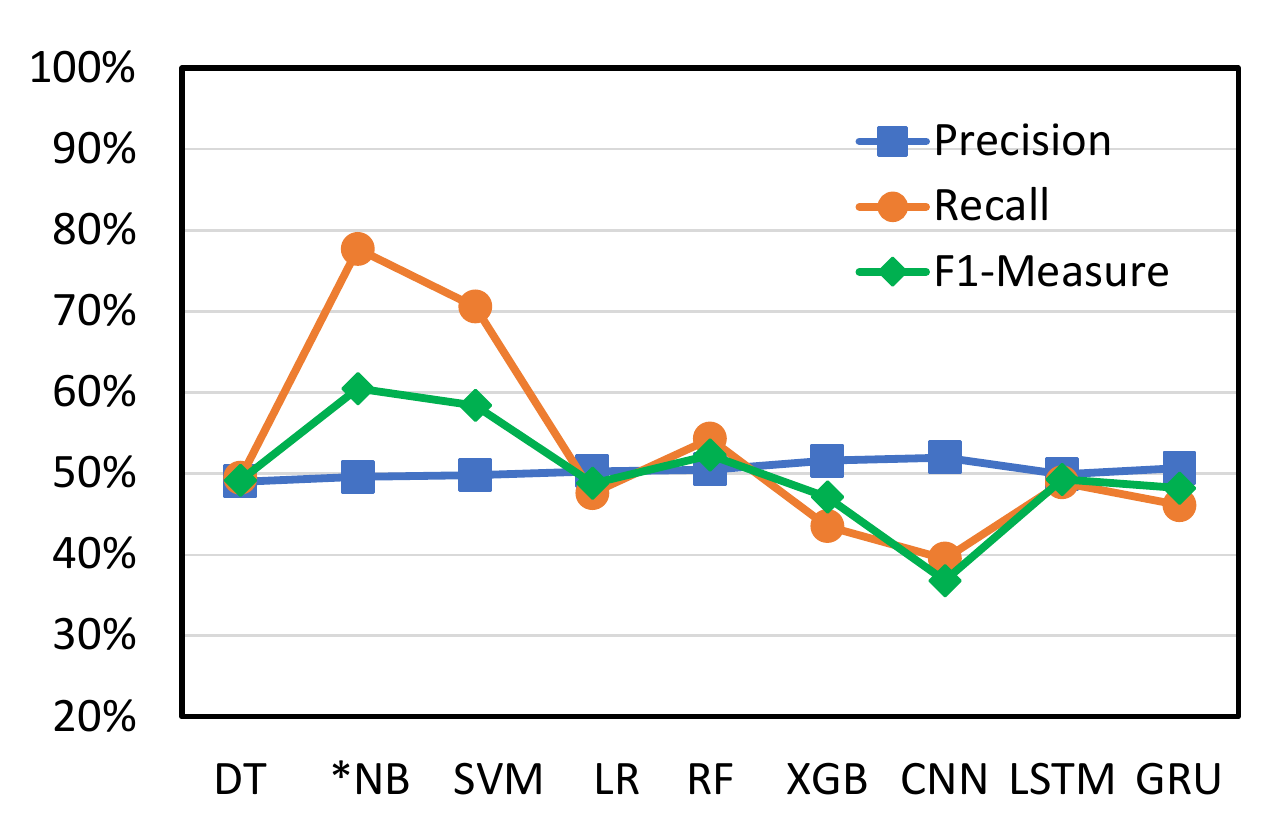}}
	\caption{Classifier performance on CV+SN, CS+LN, and CS+SN. CV: Code2Vec, CS: Code2Seq, SN: SDNE, LN: Line.}
	\label{fig:RQ2-result}
\end{figure*}

\textbf{Accuracy Evaluation (RQ2).} To evaluate the accuracy of RMove, we select three most effective embedding combinations: Code2Vec+SDNE (CV+SN), Code2Seq+Line (CS+LN), and Code2Seq+SDNE (CS+SN). Furthermore, we rank related trained models of these embedding combinations on the real-world dataset and label the most effective classifier for each combination. Therefore, we regard these combinations and related classifiers with the best performance as three variants of RMove: RMove-1, RMove-2, and RMove-3, which corresponds to Code2Vec+SDNE (CV+SN), Code2Seq+Line (CS+LN), and Code2Seq+SDNE (CS+SN) respectively. Fig.~\ref{fig:RQ2-result} presents the performance of classifiers on selected three embedding combinations. Fig.~\ref{fig:RQ2-result}.(a)-(c) presents the results of Code2Vec+SDNE (CV+SN), Code2Seq+Line (CS+LN), and Code2Seq+SDNE (CS+SN) respectively, while the most effective classifier was labeled with a * mark. Table~\ref{tab:RQ2-result} presents the accuracy of various refactoring tools on the real-world dataset including PathMove \cite{kurbatova2020recommendation}, JDeodorant \cite{tsantalis2009identification}, JMove \cite{terra2018jmove}, RMove-1, RMove-2, and RMove-3. For each row/subject, we highlight the greatest precision, recall and f1-measure results with a grey background color and a + mark. Specifically, for the row: ``avg'', we highlight the maximum average scores of precision, recall and f1-measure on studied subjects with a grey background color and a * mark. Accordingly, we label the most accurate refactoring tool with a grey background color. 

As presented in Table~\ref{tab:RQ2-result} and Fig.~\ref{fig:RQ2-result}, we observed that 1) Naive Bayes outperforms other classifiers in all of selected embedding combinations including Code2Vec+SDNE (CV+SN), Code2Seq+Line (CS+LN), and Code2Seq+SDNE (CS+SN) on the real-world dataset. Naive Bayes has the greatest recall score of all classifiers, resulting in a higher f1-measure score. This result demonstrates that Naive Bayes outperforms Random Forest (RF) and Extreme Gradient Boosting (XGB) on the real-world dataset despite having the best performance on the synthetic dataset in RQ1. One possible reason might be that Naive Bayes is more prone to generalize to other drastically different datasets in comparison to other classifiers. 2). RMove demonstrates an increase of 14\%-36\% in precision, 19\%-45\% in recall, and 27\%-44\% in f1-measure compared to stat-of-the-art tools: PathMove \cite{kurbatova2020recommendation}, JDeodorant \cite{tsantalis2009identification}, and JMove \cite{terra2018jmove} while the statical test results also show that RMove is significantly better than these three tools. Despite PathMove's high precision on 2 subjects and JMove's high recall on 1 subject, RMove outperforms other refactoring tools in precision, recall, and f1-measure on most subjects of the real-world dataset. Specially, RMove-1 presents the increase of recall scores while RMove-2 presents the improvement of precision scores. In real scenarios, software practitioners may choose the proper embedding combinations and classifiers as settings for recommending move method refactorings.

\vspace{1mm}
\noindent\fbox{
 \parbox{0.95\linewidth}{
  \textbf{Answer to RQ2}:
  RMove has an increase of 14\%-36\% in precision, 19\%-45\% in recall, and 27\%-44\% in f1-measure compared to stat-of-the-art refactoring tools.
 }
 }

\vspace{2mm}
\textbf{Usefulness Evaluation (RQ3).} To evaluate the usefulness of RMove, we conducted a user study with 15 participants to review detection results of 6 refactoring tools on the subject FreeMind in RQ2. The studied subject: FreeMind is suitable for understanding. All the participants are industrial engineers having at least 6 years of working experience, which are also not the authors of this paper. We offered the source code of FreeMind and 6 groups of detection results and each group has 5 instances, which have been sampled several times to limit the overlap of these samples. We blindly present each group of detection results in a random order to each participant to ensure they do not know which tool was developed by us. After reviewing all the groups, participants are required to evaluate each group’s performance. We further asked the participants to complete a questionnaire about refactoring tools. For each tool, we ask participants the question ``Would you apply the refactoring tool?'' and provide them with 5 ranking options including ``Definitely Not'',  ``No'', ``Maybe'', ``Yes'', and ``Absolutely Yes''.

\begin{figure} 
	\centering
	\includegraphics[scale=.45]{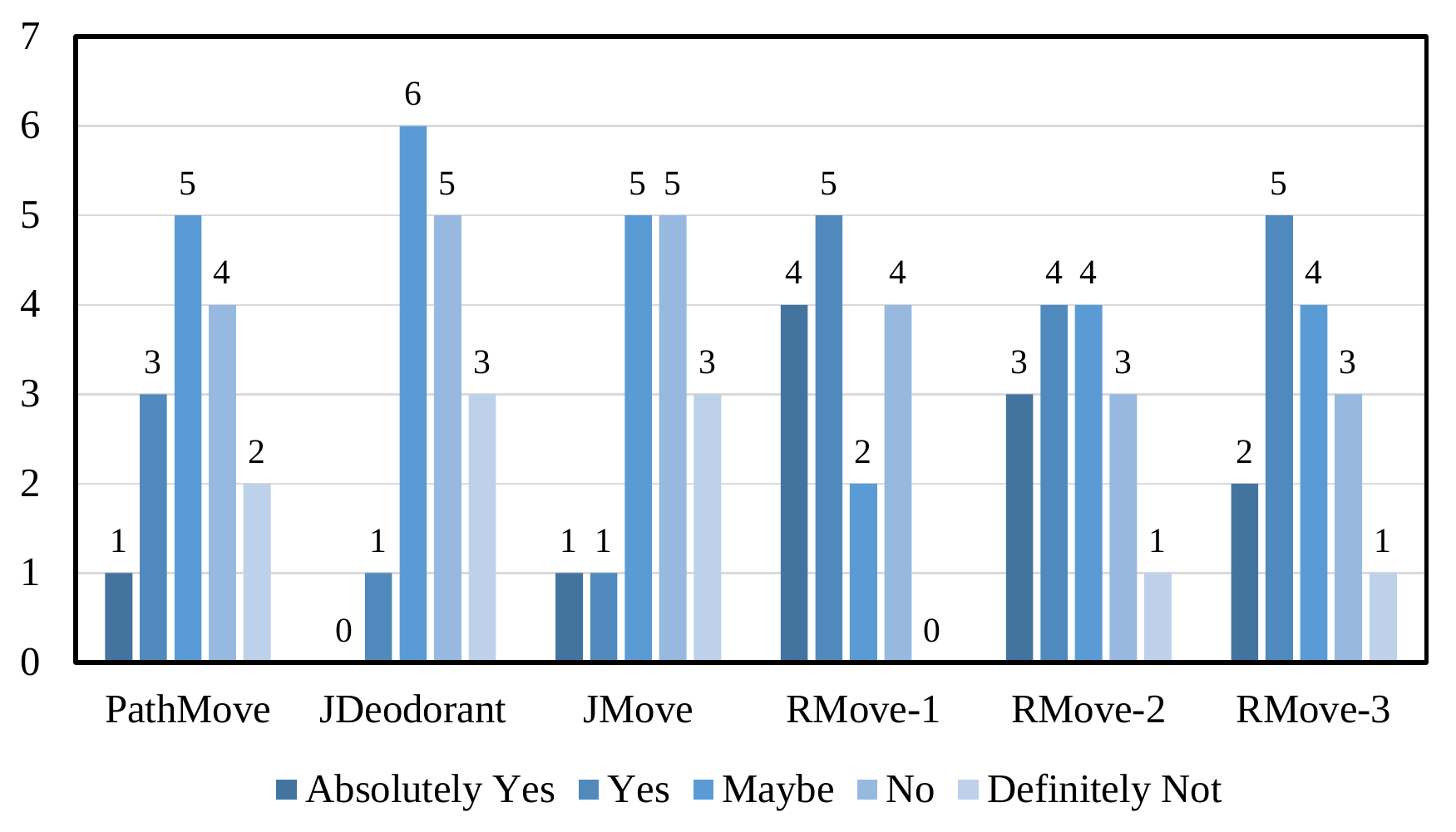}
	\caption{The distribution of participants' answers for each refactoring tool.}
	\label{fig:userstudy}
\end{figure}

Table~\ref{tab:RQ3-result} reports the results of the questionnaire of 15 participants. The first column presents the refactoring tool. The other columns present the answer of each participant for each refactoring tool. Fig.~\ref{fig:userstudy} presents the distribution of participants' answers for each refactoring tool. We observed that 1) Generally, the majority of participants believe that RMove is more helpful compared with the results of state-of-the-art refactoring tools. 7-9 participants choose ``Absolutely Yes'' or ``Yes'' for RMove while 1-4 participants support other tools. 2) More participants believe that state-of-the-art refactoring tools are not very helpful. 6-8 participants choose ``Definitely Not'' or ``No'' for these refactoring tools. 3) More participants hesitate to use state-of-the-art refactoring tools. 5-6 participants choose ``Maybe'' for these refactoring tools. This result indicates that RMove is more likely than state-of-the-art refactoring tools to be embraced by software practitioners. 

\vspace{1mm}
\noindent\fbox{
 \parbox{0.95\linewidth}{
  \textbf{Answer to RQ3}:
  RMove is more useful than other refactoring tools in recommendation of move method refactoring opportunities for most participants.
 }
}

\begin{table*}[]
\centering
\caption{Participants' answers to the question ``Would you apply the refactoring tool?''. AY: Absolutely Yes, Y: Yes, M: Maybe, N: No, DN: Definitely Not.}
\small
\label{tab:RQ3-result}
\begin{tabular}{lccccccccccccccc}
\hline
\multicolumn{1}{l}{Tool} & P1 & P2 & P3 & P4 & P5 & P6 & P7 & P8 & P9 & P10 & P11 & P12 & P13 & P14 & P15 \\ \hline
PathMove & AY & N & Y & M & N & M & M & Y & DN & M & N & N & DN & Y & M \\
JDeodorant & M & M & DN & N & N & M & N & N & N & DN & M & M & DN & M & Y \\
JMove & M & Y & M & N & M & N & M & DN & M & DN & N & AY & N & DN & N \\
RMove-1 & AY & Y & N & AY & Y & AY & Y & N & Y & M & AY & N & M & Y & N \\
RMove-2 & Y & M & N & AY & M & Y & Y & N & Y & DN & AY & N & M & AY & M \\
RMove-3 & M & N & M & AY & Y & Y & AY & M & Y & M & Y & N & DN & Y & N \\ \hline
\end{tabular}
\end{table*}
\setlength{\belowcaptionskip}{-0.6cm}

\section{Discussion}
\label{sec:discussion}

In this section, we discuss the limitation, threats to validity, and applications of our approach.

\noindent \textbf{Threats. }Our research has the following threats: The first threat comes from data used in our evaluation. We only evaluate our approach in a part of selected open-source projects. It is still unclear whether our approach will generalize to closed source industrial projects and open-source projects from other communities. Replicating our approach on more datasets is our ongoing work. The second threat comes from the data quality of our training data. The collected data may contain noise which may lead to the bias of model training. To limit this threat, we employ the popular open-source projects, which are maintained by active communities containing less noise. We further manually check each item of training data to improve its quality. The third threat comes from the feasible evaluation of our recommendation results. We employed 15 participants to manually check the preconditions of our move method recommendation results. In our future work, we will further leverage automatic precondition validation techniques to ensure the refactoring recommendation solutions are actually applicable. The fourth threat comes from the run-time evaluation of our approach. We only test the run-time of RMove on real-world dataset. On these small subjects, RMove takes a relatively long time to process data and train model. However, once these phrases are completed, RMove can return refactoring recommendation results rapidly. Table~\ref{tab:run-time} presents the run-time performance of RMove on the real-world dataset. On average, RMove responds within 2 seconds. Analyzing the run-time performance of our approach on large scale subjects is our ongoing work.

\noindent \textbf{Limitations.} Our research has several limitations, which are illustrated as follows: First, machine learning techniques are significantly influenced by their hyper-parameters. To limit this threat, we employ grid search to explore suitable hyper-parameters of classifiers. However, we follow most of the default hyper-parameters of embedding techniques used in previous study \cite{alon2019code2vec,alon2018code2seq,qu2021evaluating}, which may cause sub-optimal results. In our future work, we will further explore the impact of hyper-parameters of embedding techniques including code embedding and network embedding. Second, we employ the implementation of embedding techniques in previous work \cite{alon2019code2vec,alon2018code2seq,qu2021evaluating}, which may produce wrong results. To reduce this threat, we reported our found defects. If they are not fixed, we tried to fix them by ourselves. This can be further reduced by using more advanced tools. Third, we only focus on calling relationships between methods to generate structural representation. We did not consider fine-grained information of method calling behaviours as features in our approach, like various types of callees and calling frequencies. We believe these fine-grained features can be potentially useful and will further leverage them to improve our approach in future work.

\begin{table*}[htbp]
\centering
\caption{The run-time performace of RMove.}
\label{tab:run-time}
\small
\begin{tabular}{lccccccccc}
\hline
Tool & Weka & Ant & FreeCol & JMeter & FreeMind & JTOpen & DrJava & Maven & Avg \\ \hline
Rmove-1 & 1.45s & 1.86s & 1.97s & 1.46s & 1.37s & 1.52s & 1.30s & 1.47s & 1.55s \\
Rmove-2 & 1.06s & 1.44s & 1.51s & 1.06s & 0.93s & 1.08s & 0.90s & 1.07s & 1.13s \\
Rmove-3 & 1.61s & 1.96s & 2.06s & 1.53s & 1.47s & 1.62s & 1.47s & 1.55s & 1.66s \\ \hline
\end{tabular}
\end{table*}

\noindent \textbf{Applications.} Our research can be further extended in several directions, which are listed as follows: First, our approach uses the synthetic dataset to train the classification model and some results are promising. Despite its small size, the synthetic dataset is of great quality. This implies that in refactoring recommendation tasks, data quality is more significant than data quantity. In future work, high-quality synthetic data could be leveraged to improve the performance of refactoring recommendation techniques. Second, our results indicate that various embedding combinations have drastically different effects on the refactoring recommendation results. Furthermore, the performance of classifiers also varies depending on the embedding combination. This opens up the possibility of using the ensemble technique to combine several variants of our approach implemented with various embedding combinations and classifiers. Designing the proper ensemble technique to improve the refactoring recommendation in future work is deserved. Third, our results demonstrate that improving the move method refactoring recommendation using structural and semantic representations of code snippets is feasible. It implies that the combination of structural and semantic representations could help with further additional feature envy refactoring techniques such as move attribute, move class and move package, which are all related to the inappropriate placement of code entities. Our future work will focus on the existing refactoring recommendation technique improving by exploiting both the structural and semantic representation of code snippets.

\section{Related Work}
\label{sec:relatedwork}

\setlength{\belowcaptionskip}{-0.2cm}
Over the past decades, numerous tools have been designed to automatically suggest Move Method refactoring. In this section, we provide an overview of existing approaches and categorize them broadly into two groups:

\textbf{Non-machine-learning-based approaches}. The most representative work in this area is JDeodorant, which was introduced by Tsantalis et al. \cite{tsantalis2009identification}. JDeodorant supports the detection of several types of code smells, including Feature Envy, Long Method, and God Class, and recommend appropriate refactoring suggestions to remove them. JDeodorant detects Feature Envy based on the following rule: a method should be moved when this method accesses more entities in other classes than entities in its class. To ensure the correctness of Move Method refactoring suggestions, JDeodorant also employs a set of preconditions and automatically verifies them. Terra et al. \cite{terra2018jmove} introduced a Move Method refactoring tool: JMove based on the similarity between dependency sets. JMove considers dependencies such as method calls, field accesses, return types, etc. JMove detects Feature Envy according to the dependencies established by the method. The results show that JMove outperforms JDeodorant in accuracy and efficiency, especially for large methods. Liu et al. \cite{liu2016domino} introduced a Move Method refactoring tool: Domino based on the movement of other methods. Domino detects Feature Envy based on a heuristic that similar methods should be moved together. When a method is moved, Domino explores possible methods and suggests to move. Ujihara et al. \cite{ujihara2017c} introduced a Move Method refactoring tool: C-JRefRec based on static program analysis. C-JRefRec detects Feature Envy by computing the semantic similarity between the method and target class with TF-IDF vectors. Bavota et al. \cite{bavota2013methodbook} introduced a Move Method refactoring tool: MethodBook using Relational Topic Model (RTM). MethodBook considers both method calls and textual information, including identifiers and comments, to support the detection of Feature Envy. These proposed approaches are based on software metrics, which require expert knowledge to define rules and thresholds, whereas our approach employs embedding techniques to learn features from code snippets automatically.


\textbf{Machine learning-based approaches.} Liu et al. \cite{liu2019deep} proposed a deep learning-based approach to detect several code smells including Feature Envy, Long Method, and Large Class. They first extract textual information such as identifiers and generate its representations with the word2vec technique \cite{mikolov2013efficient}. Then, they further calculate the distance between representations and use Convolutional Neural Network (CNN) model to identify Feature Envy. Hadj-Kacem et al. \cite{hadj2019deep} also proposed a deep learning-based approach to detect several code smells including Feature Envy, Long Method, and Blob. They parse the source code into Abstract Syntax Trees (AST) and generate its representation using the Variational Auto-Encoder (VAE) \cite{palomba2018diffuseness}. They further use the Linear Regress classifier to detect Feature Envy. Sharma et al. \cite{sharma2019feasibility} systematically compare the detection of code smells including Complex Method, Magic Number, Empty Catch Block, and Multifaceted Abstraction with deep learning models. The results present that Recurrent Neural Network performs the best. Bryksin et al. \cite{bryksin2018automatic} introduced a Move Method refactoring tool: ArchReload using clustering ensemble techniques. ArchReload combines the detection results of several heuristic-based approaches, such as ARI \cite{marian2012using}, HAC \cite{marian2012study}, and CCDA \cite{pan2013refactoring}. Barbez et al. \cite{barbez2020machine} also introduced a Move Method refactoring tool with classifier ensemble techniques. They first collect software metrics and further employ these metrics to train machine learning classifiers and ensemble them. These approaches extract identifiers from code snippets and characterize features using word2vec techniques, while our approach uses graph embedding techniques and code embedding techniques to better capture the structure and semantic properties of code snippets.


\section{Conclusion}
\label{sec:conclusion}

In this paper, we proposed an approach to recommend Move Method refactoring named RMove by automatically learning both structural and semantic representation from code snippets. We first extract method structural and semantic information from the dataset. Next, we create the structural and semantic representation, and further concatenate these representations. Finally, we train the machine learning classifier to guide the movement of method to suitable class. The results show that our approach outperforms three state-of-the-art refactoring tools including PathMove, JDeodorant, and JMove.

\section*{acknowledgement}

This work was supported by National Natural Science Foundation of China (61902288, 61972300, U21B2015), Strategic Priority Research Program of Chinese Academy of Science(XDC05040100), National Key Research and Development Program of China under Grant (2019YFB1406404), Fundamental Research Funds for the Central Universities (XJS220311).

\bibliographystyle{IEEEtran}
\bibliography{ref}

\end{document}